\documentclass[10pt]{iopart}
\usepackage{longtable}
\usepackage{graphicx}
\usepackage{times}
\usepackage{amssymb}
\newcommand{\beq}{\begin{equation}}
\newcommand{\eeq}{\end{equation}}
\newcommand{\bea}{\begin{eqnarray}}
\newcommand{\eea}{\end{eqnarray}}

\begin{document}

\title[]{Squeezed Light for the Interferometric Detection of High Frequency Gravitational Waves}

\author{R.~Schnabel\dag
\footnote[3]{To whom correspondence should be addressed
(Roman.Schnabel@aei.mpg.de)} , J.~Harms\dag, K.~A.~Strain\ddag,
and K.~Danzmann\dag}

\address{\dag\ Max-Planck-Institut f\"ur Gravitationsphysik (Albert-Einstein-Institut) and\\
Institut f\"ur Atom- und Molek\"ulphysik, Universit\"at
Hannover, 30167 Hannover, Germany}
\address{\ddag\ Department of Physics and Astronomy, University of Glasgow, Glasgow G12 8QQ, UK}

\begin{abstract}
The quantum noise of the light field is a fundamental noise source
in interferometric gravitational wave detectors. Injected squeezed
light is capable of reducing the quantum noise contribution to the
detector noise floor to values that surpass the so-called
Standard-Quantum-Limit (SQL). In particular, squeezed light is
useful for the detection of gravitational waves at high
frequencies where interferometers are typically shot-noise
limited, although the SQL might not be beaten in this case. We
theoretically analyze the quantum noise of the signal-recycled
laser interferometric gravitational-wave detector GEO\,600 with
additional input and output optics, namely frequency-dependent
squeezing of the vacuum state of light entering the dark port and
frequency-dependent homodyne detection. We focus on the frequency
range between 1 kHz and 10 kHz, where, although signal recycled,
the detector is still shot-noise limited. It is found that the
GEO\,600 detector with present design parameters will benefit from
frequency dependent squeezed light. Assuming a squeezing strength
of $-6$~dB in quantum noise variance, the interferometer will
become thermal noise limited up to 4 kHz without further reduction
of bandwidth. At higher frequencies the linear noise spectral
density of GEO\,600 will still be dominated by shot-noise and
improved by a factor of $\,10^{6\rm{dB}/20\rm{dB}} \approx 2$
according to the squeezing strength assumed. The interferometer
might reach a strain sensitivity of $6\times10^{-23}$ above 1 kHz
(tunable) with a bandwidth of around 350~Hz. We propose a scheme
to implement the desired frequency dependent squeezing by
introducing an additional optical component to GEO\,600s
signal-recycling cavity.
\end{abstract}




\section{Introduction}

It was first proposed by Caves~\cite{Cav81} that squeezed light
injected into the dark port of a gravitational wave (GW)
interferometer can be employed to reduce the high laser power
requirements. Later Unruh~\cite{Unruh82} and
others~\cite{GLe87,JRe90,PCW93,KLMTV01} have found and proven in
different ways that squeezed light with a frequency dependent
orientation of the squeezing ellipse can reduce the quantum noise
over the complete spectrum. Therefore quantum noise spectral
densities below the so-called standard quantum limit (SQL) are
possible. In all cases interferometer topologies without
signal-recycling were considered. Chickarmane {\it et al.}
\cite{CDh96,CDRGBM98} investigated the signal-recycled
interferometer with Fabry-Perot arm cavities and injected squeezed
light. Radiation pressure noise was not taken into account.
Recently Harms {\it et al.} \cite{HCCFVDS03} have shown that high
power signal-recycled interferometers will benefit from squeezed
light similarly to conventional interferometers.

\begin{figure}[ht!]
\centerline{\includegraphics[angle=0,width=6cm]{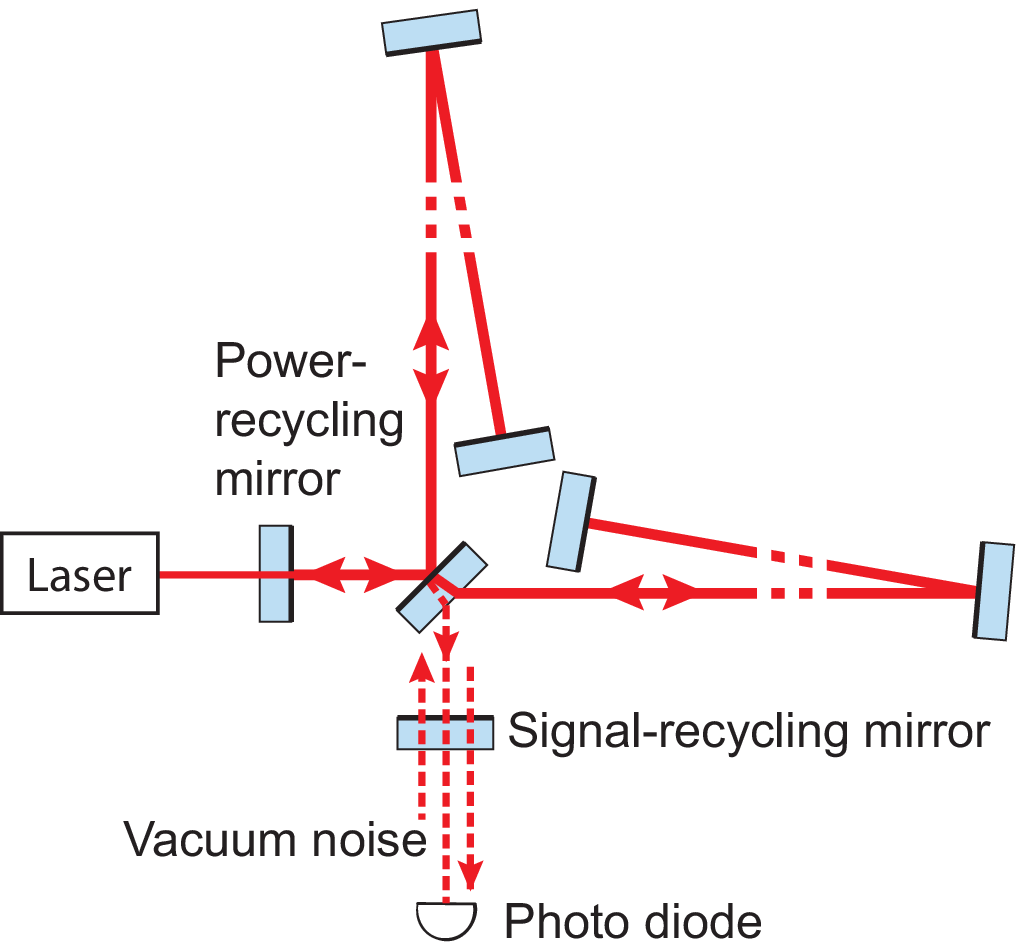}\hspace{10mm}\vspace{9mm}\includegraphics[angle=0,width=5cm]{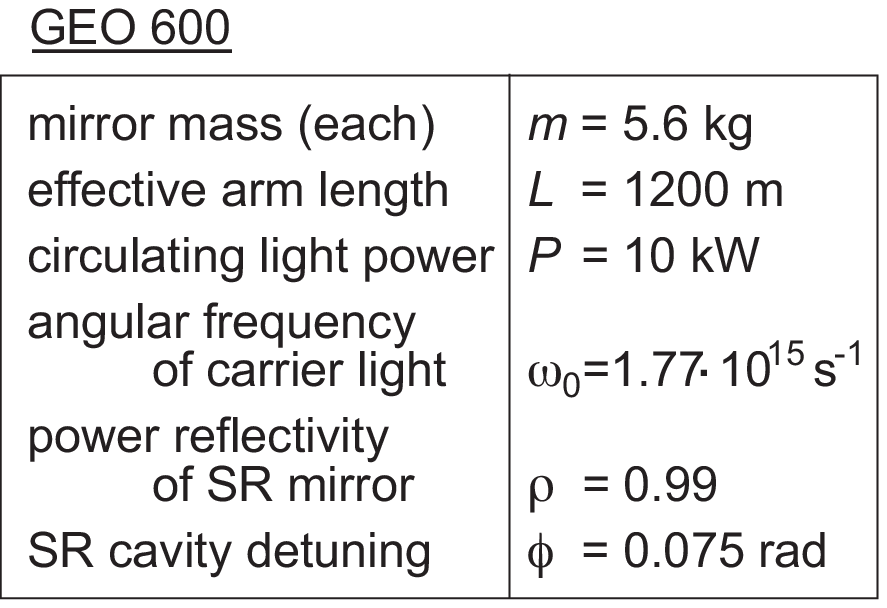}}
  \vspace{-10mm}
  \caption{GEO\,600 is a dual-recycled Michelson interferometer with a power-recycling mirror in the bright
  port that enhances the light power within the Michelson arms and a signal-recycling mirror in the dark port
  that can be tuned on a specific signal frequency. Since the arms are folded once, then the effective arm length
  is doubled to 1200~m. Vacuum noise enters the interferometer from the dark port and is back reflected onto the
  photo diode. The table provides technical data and parameter values of GEO\,600 which where used to calculate
  the noise spectral densities in Figs.~\ref{noise1} and \ref{noise3}a; SR: signal-recycling.}
  \label{GEO600}
\end{figure}
GEO\,600 \cite{geo02} is the only first-generation detector that
not only uses power recycling \cite{DHKHFMW83pr}, but also
includes the more advanced technique of {\it signal recycling}.
The original idea of the signal-recycling (SR) was due to Meers
\cite{Mee88}, who proposed its use for {\it dual-recycling}, which
is the combination of power- and signal-recycling. With signal
recycling, part of the GW induced signal light is retro-reflected
at the dark port and back into the interferometer, establishing an
additional cavity which can be set to resonate at a desired
gravitational-wave frequency (Fig.~\ref{GEO600}). Signal recycling
leads to a well known (optical) resonance structure in the
interferometer's noise curve. This resonance can already beat the
standard quantum limit even in the absence of squeezed light input
\cite{BCh01b}. Buonanno and Chen have also predicted a second,
opto-mechanical resonance in signal-recycled interferometers,
around which the interferometer gains sensitivity, and which can
also beat the standard quantum
limit~~\cite{BCh01b,BCh01a,BCh02a,BCh03a}.
%

In this paper, we investigate the benefit of injecting squeezed
vacuum with frequency-dependent squeezing angle into GEO\,600s
dark port. We focus on the frequency range between 1~kHz and
10~kHz where the interferometer is expected to be shot-noise
limited. We use the two-photon input-output formalism of quantum
optics \cite{CSc85} which has also been used by Harms {\it et al.}
in \cite{HCCFVDS03}. Our results are presented in terms of noise
spectral densities and are compared with the thermal noise
expected for GEO\,600. Several different types of astrophysical
sources have been predicted, that might radiate gravitational
waves in the kHz spectrum. Generally speaking, small objects of 1
to 10 solar masses and of about 10~km diameters are considered.
Due to theoretical models, neutron star bar-mode instabilities,
black hole ringing, merging neutron star binaries  \cite{FHH02}
and other sources are discussed. For an overview we refer to the
article by Kokkotas in this issue.

\section{Spectral noise densities of GEO\,600}
\label{current}

GEO\,600 is, as already mentioned above, a dual-recycled
interferometer with folded interferometer arms. Recently the
GEO\,600 interferometer in Ruthe near Hannover has been completed
by the  implementation of the signal-recycling mirror. Relevant
technical parameters of GEO\,600 are summarized in
Fig.~\ref{GEO600}.

\begin{figure}[ht!]
\centerline{\includegraphics[width=6.5cm]{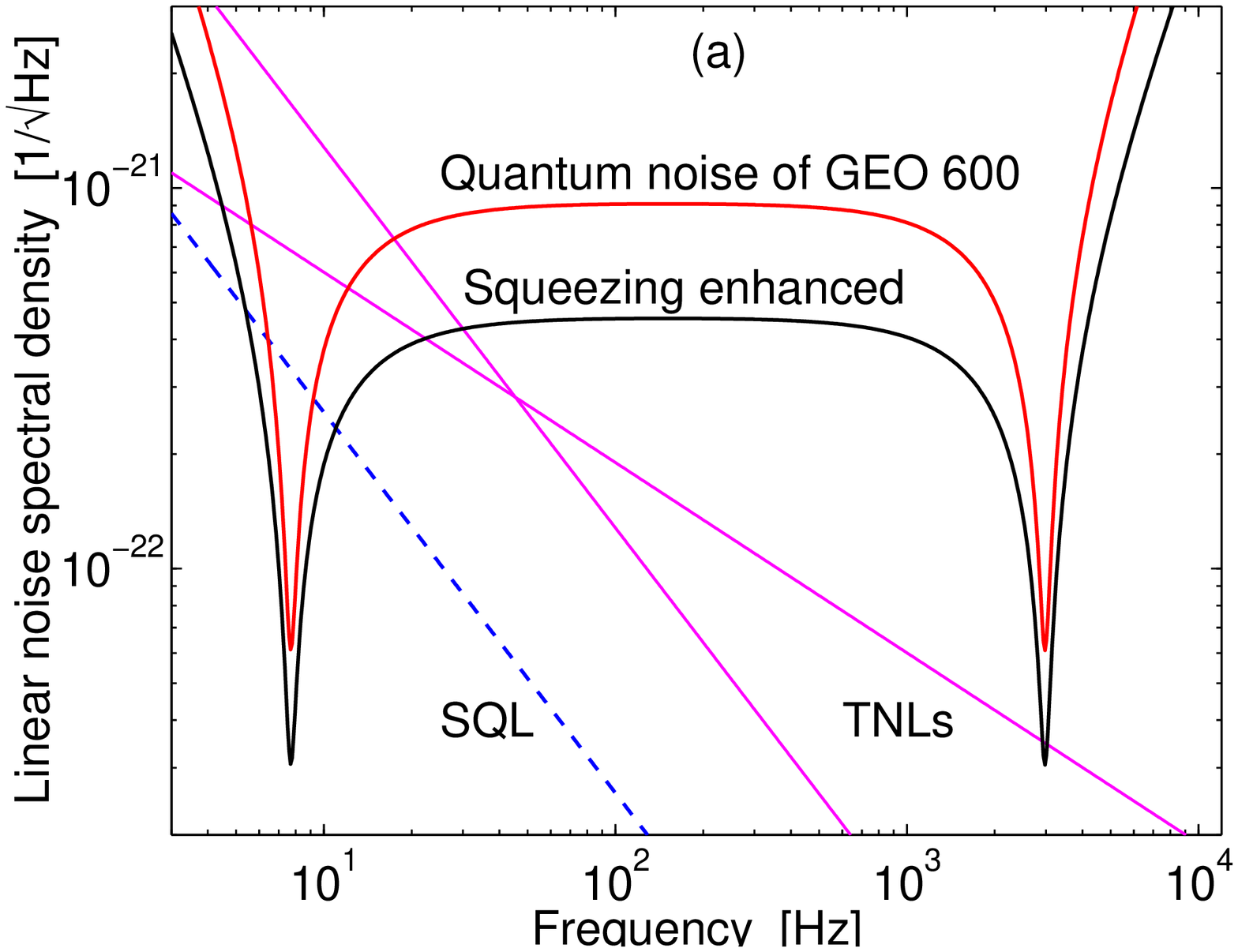}\,\,\,\includegraphics[width=6.15cm]{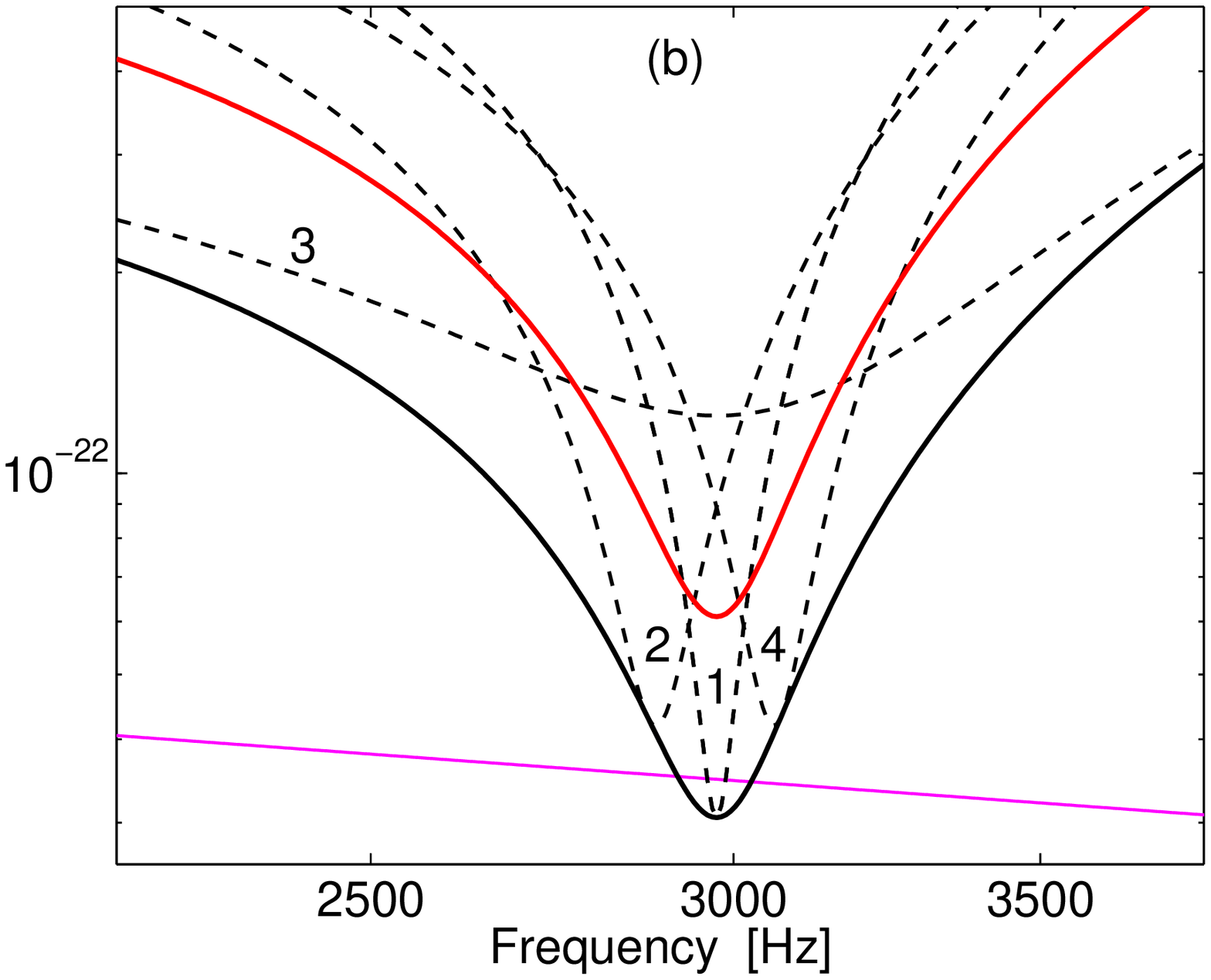}}
  \vspace{0mm}
  \caption{a) The linear quantum noise spectral density can be reduced by
  a factor of $\,10^{x/20\rm{dB}} = \rm{e}^{-r}$, where x is the relative
  squeezing of the variance and r is the so-called squeezing parameter.
  Here we chose $x=6~\rm{dB}$. The amplitude quadrature is shown with and
  without optimally frequency dependent squeezed light again in comparison
  with the standard quantum limit (SQL) and thermal noise limits (TNLs).\\
b) Zoom into the optical resonance. The dashed lines give the
results of applying frequency independent squeezed light, where
the squeezing is chosen to be in amplitude (curve 1), at
$45^\circ$ (2), in phase (3) and at $-45^\circ$ (4). The lower
curve is again the optimum achievable, when applying the correct
squeezing angle at every frequency.}
  \label{noise1}
\end{figure}

Here we are interested in the optical noise of GEO\,600 above
1~kHz. Quantum noise and thermal noise have to be considered.
Generally, the optical noise in an interferometer can be expressed
in terms of the (single-sided) noise spectral density of the
output field normalized by the transfer function of the signal.
Fig.~\ref{noise1}a shows expected linear noise spectral densities
(square root of the noise spectral densities) of the GEO\,600
interferometer. The upper grey curved line in Fig.~\ref{noise1}a
represents the quantum noise in the amplitude quadrature. The
straight solid lines represents the internal ($\propto
\sqrt{1/f}$) and the photo-refractive thermal noise limits
($\propto 1/f$). GEO\,600s standard quantum limit (SQL) is given
as a reference (straight dashed line):\vspace{-2mm}
\beq \sqrt{h_{\rm SQL}} = \sqrt{\frac{20\hbar}{m\Omega^2 L^2}} \,.
\eeq

We have used the two-photon {\it input-output relation} of quantum
optics \cite{CSc85}, which maps the numerous input fields
$\overline{\mathbf{i}}_n$ and the gravitational-wave signal
$h=\Delta L/L$ onto the detected output field
$\overline{\mathbf{o}}$. $\overline{\mathbf{i}}$ and
$\overline{\mathbf{o}}$ are two-dimensional vectors, where the
first components represent the fields amplitude quadrature and the
second ones the phase quadrature. The general expression for the
resulting quantum noise spectral density $S_h$ can be found in
\cite{HCCFVDS03}. We note that optical losses and radiation
pressure noise at the beam-splitter is not considered here.
According to the table in Fig.~\ref{GEO600} we chose the
signal-recycling cavity to be detuned from the carrier frequency
by 0.075 rad. The SR cavity is therefore resonant for 3~kHz
gravitational wave signals. It can be seen in Fig.~\ref{noise1}a
that at this {\it optical} resonance GEO\,600 is expected to be
quantum noise (shot-noise) limited. The second resonance at lower
frequencies stems from the classical opto-mechanical coupling of
the light field with the anti-symmetric mode of the otherwise free
mirrors~\cite{BCh02a}: in detuned signal-recycling schemes, the
phase-modulation sidebands induced by a gravitational wave are
partly converted into amplitude modulation, which beat with the
carrier field, producing a motion-dependent force and acting back
on the test masses. This {\it opto-mechanical} resonance is likely
to be buried by thermal noise which is also expected for GEO\,600.
Only above 1\,kHz will squeezed light injected into the dark port
of the interferometer be able to reduce the overall noise-floor.

For all noise spectra in this paper we assume homodyne detection
at the dark port of the interferometer. The detected quadrature of
angle $\zeta$ is assumed to be the amplitude quadrature in respect
to the carrier field inside the interferometer ($\zeta=0$). An
appropriate local oscillator can be provided by differential loss
in the two interferometer arms. In the case of no differential
loss, but while locking the interferometer slightly offset to the
dark fringe, the phase quadrature will be detected which is the
upper dashed line in Fig.~\ref{noise3}a. Apart from more quantum
noise at intermediate frequencies, all results from this paper
remain valid for homodyne detection of the phase quadrature or of
any other quadrature of angle $\zeta$. The quantum noise in a
heterodyne readout scheme has also been analyzed and tends towards
higher noise floors \cite{BCM03}.

\section{Squeezed light enhanced GEO\,600}
\label{SI}

Generally, quantum noise can be understood in terms of vacuum
fluctuations. Since the interferometer is operated at the dark
fringe, vacuum fluctuations impinging on the photo diode at the
interferometers dark port stem from vacuum fluctuations that have
been reflected back. Indeed the beat between these vacuum
fluctuations and the carrier laser light entering the bright port
excites the interferometers anti-symmetric mode of motion. On the
other hand, vacuum noise entering the bright port excites the
irrelevant symmetric mode of motion and does not effect the
quantum noise of the GW detector. For this reason, squeezed light
is planned to be injected into the interferometers dark port.
Recently, a scheme using a Faraday rotator has been experimentally
demonstrated on a table-top power-recycled
Michelson-interferometer \cite{KSMBL02}. Squeezed light is
currently most efficiently generated in optical parametric
amplifiers (OPAs). At a laser wavelength of 1064~nm, MgO:LiNbO$_3$
crystals are heated to about $107^\circ \rm C$ to meet the phase
matching condition between the fundamental and the green second
harmonic pump frequency. The relative phase shift of both
frequencies can be locked by a servo loop and determines what
quadrature of the infrared OPA output will be squeezed. The
orthogonal quadrature will then be anti-squeezed at least by the
same factor due to the {\it Heisenberg uncertainty principle}. A
reduction of quantum noise variances of up to 7~dB has been
reported at sideband frequencies of a few MHz with squeezing
bandwidths of also a few MHz \cite{SLMS98}, \cite{LRBMBG99}. For
the use in GW interferometers broadband squeezed light at GW
frequencies needs to be generated with a frequency dependent
rotation of the squeezed quadrature. The lowest frequency that
squeezed light has been demonstrated at so far is at 220~kHz
\cite{BSTBL02}. Frequency dependent squeezing has not been
demonstrated yet.

Fig.~\ref{noise1}a shows the reduction of quantum noise in the
amplitude quadrature expected for GEO\,600 if a non-classical
vacuum state which is optimally frequency dependently squeezed in
variance by -6~dB throughout the complete spectrum from 3 to
10000~Hz is injected into the dark port. In Fig.~\ref{noise1}b we
zoom into the optical resonance. The dashed lines give the results
of applying frequency independent squeezed light, where the
squeezing is chosen to be in amplitude, at $45^\circ$, in phase
and at $-45^\circ$, respectively. It is well-known that frequency
dependent squeezed light is crucial for a GW interferometer to
benefit from non-classical quantum noise reduction. We therefore
focus now on the issue how to implement appropriate filter
cavities that convert frequency independent into frequency
dependent squeezed vacuum states of light.

\begin{figure}[ht!]
\centerline{\includegraphics[angle=0,width=5.5cm]{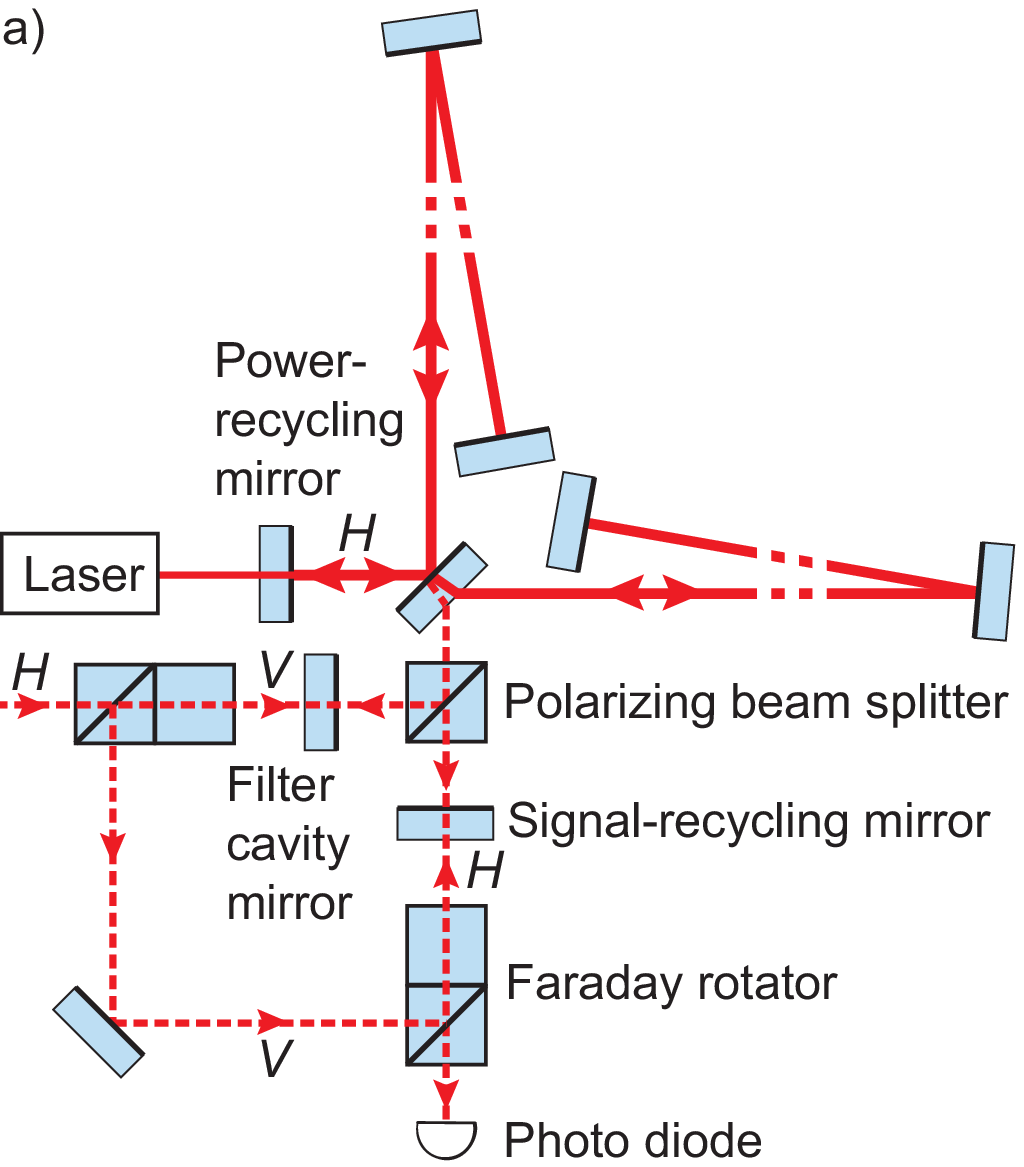}\hspace{7mm}\includegraphics[width=6.5cm]{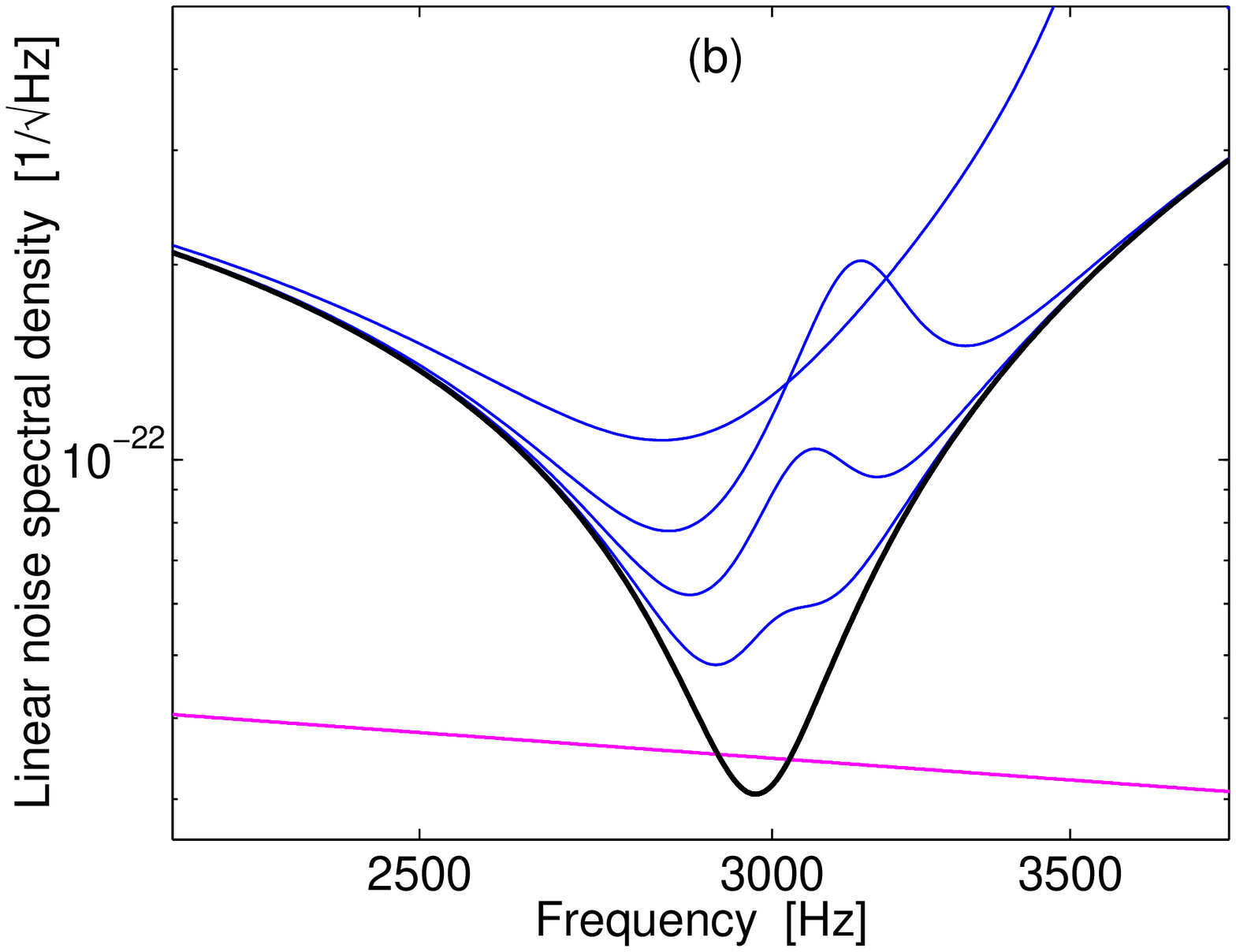}}
  \vspace{0mm}
  \caption{a) Proposed scheme to cancel the phase shift on frequency
  independent squeezed light. {\it H} and {\it V} label the horizontal
  and vertical polarization modes respectively.\\
b) The detuning of the filter cavity is varied from its here
optimum value of $\phi_{\rm FP} = -0.075\,\rm{rad}$ by -0.016,
-0.004, -0.002 and -0.001 rad (from top to bottom).}
  \label{GEO600upgrade}
  \label{noise4}
\end{figure}

As said above, squeezed vacuum can be generated with a variable
but frequency-independent squeezing angle $\lambda$ (see for
example \cite{BISM98}). A frequency-dependent squeezing angle can
then be obtained by subsequently filtering the initial squeezed
light through detuned Fabry-Perot (FP) cavities, as proposed by
Kimble et al.~\cite{KLMTV01} for so-called quantum-non-demolition
interferometers. Harms {\it et al.} found that for signal-recycled
GW interferometers, two filter cavities are necessary to cancel
rotations due to the SR cavity as well as rotation due to the
radiation pressure force if the readout quadrature is frequency
independent \cite{HCCFVDS03}. Here we consider an optical
resonance which is well separated from the opto-mechanical one.
Therefore the radiation pressure noise need not to be taken into
account and the quantum noise in Fig.~\ref{noise1}a above 1~kHz is
solely due to shot noise. The rotation due to the SR cavity is
therefore the only one which needs to be cancelled, again if the
readout quadrature is frequency independent. As an example, an
appropriate single filter cavity, labeled by FP, could match the
SR cavity parameters apart from a negative detuning from the
carrier light $\phi$:
\beq
\Omega^{\rm filter\,res} = \frac{\phi_{\rm FP} c}{L_{\rm FP}} - \rm{i} \frac{c \tau_{\rm FP}^2}{4 L_{\rm FP}}=- \frac{\phi c}{L} - \rm{i} \frac{c \tau^2}{4 L}\,.
\eeq
Here $\Omega^{\rm filter\,res}$ is the filter cavity resonant
frequency, $L$ is the cavity length and $\tau$ the transmissivity
of the cavity input mirror. When detecting the quadrature at angle
$\zeta$, the required initial (frequency independent) squeezing
angle $\lambda_0$ is found to be
\beq \label{initlambdaGEO} \lambda_0 = \zeta -\pi/2\,, \eeq
which puts the minor axis of the noise ellipse onto the $\zeta$
quadrature. Note, here $\lambda_0 =0$ defines squeezing of the
phase quadrature, according to definitions in Eq.~(12) of
Ref.~\cite{HCCFVDS03}. If for example the amplitude quadrature of
the light field at the dark port is detected, then, at the SR
cavity resonant frequency, the amplitude quadrature needs to be
squeezed. The phase reference is set by the carrier light field
inside the Michelson interferometer. For a more general
consideration in case of more filter cavities required we refer to
the work of Purdue and Chen in Appendix A of Ref.~\cite{PCh02} and
to the work of Harms {\it et al.} \cite{HCCFVDS03}.

In Fig.~\ref{GEO600upgrade}a we propose a scheme that implements
the desired single filter cavity. Lets assume that the laser light
inside the interferometer is horizontally polarized. A polarizing
beam splitter is placed between the SR mirror and the
interferometers beam splitter. A mirror matching the SR mirror is
added to form an additional cavity, now for the vertical mode.
This new cavity can be detuned independently from the SR cavity
and can thereby provide the desired filter cavity. If $\phi_{\rm
FP}=-\phi_{}$ then the phase shift on squeezed vacuum noise that
is subsequently reflected from both cavities cancels to zero. This
is shown in Fig.~\ref{noise4}b where the detuning of the filter
cavity is varied at the fixed detuning of the SR cavity.

\section{Variational output}
\label{VO}

The quantum noise spectral density of interferometers can benefit
simultaneously and independently from both frequency-dependent
squeezed light input, and frequency-dependent homodyne read-out
(variational output). This has been shown for conventional
interferometers without signal-recycling by Kimble {\it et al.}\
\cite{KLMTV01} and for optical-spring signal-recycled
interferometers by Harms {\it et al.}\ \cite{HCCFVDS03}. Both the
squeezed input and the variational output schemes require
additional optics and rotating filter cavities.

\begin{figure}[h!!]
\centerline{\includegraphics[width=6.5cm]{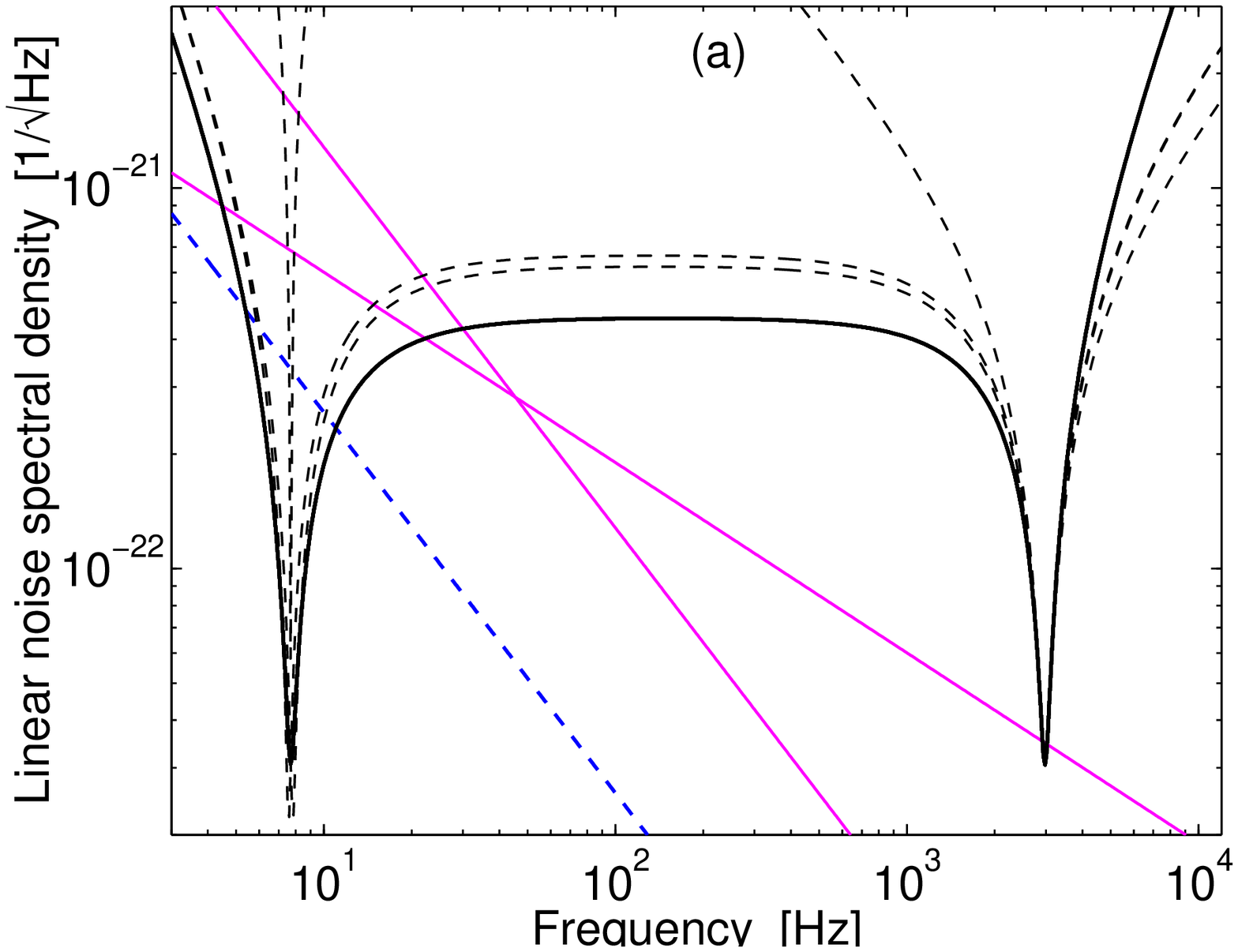}\,\,\includegraphics[width=6.15cm]{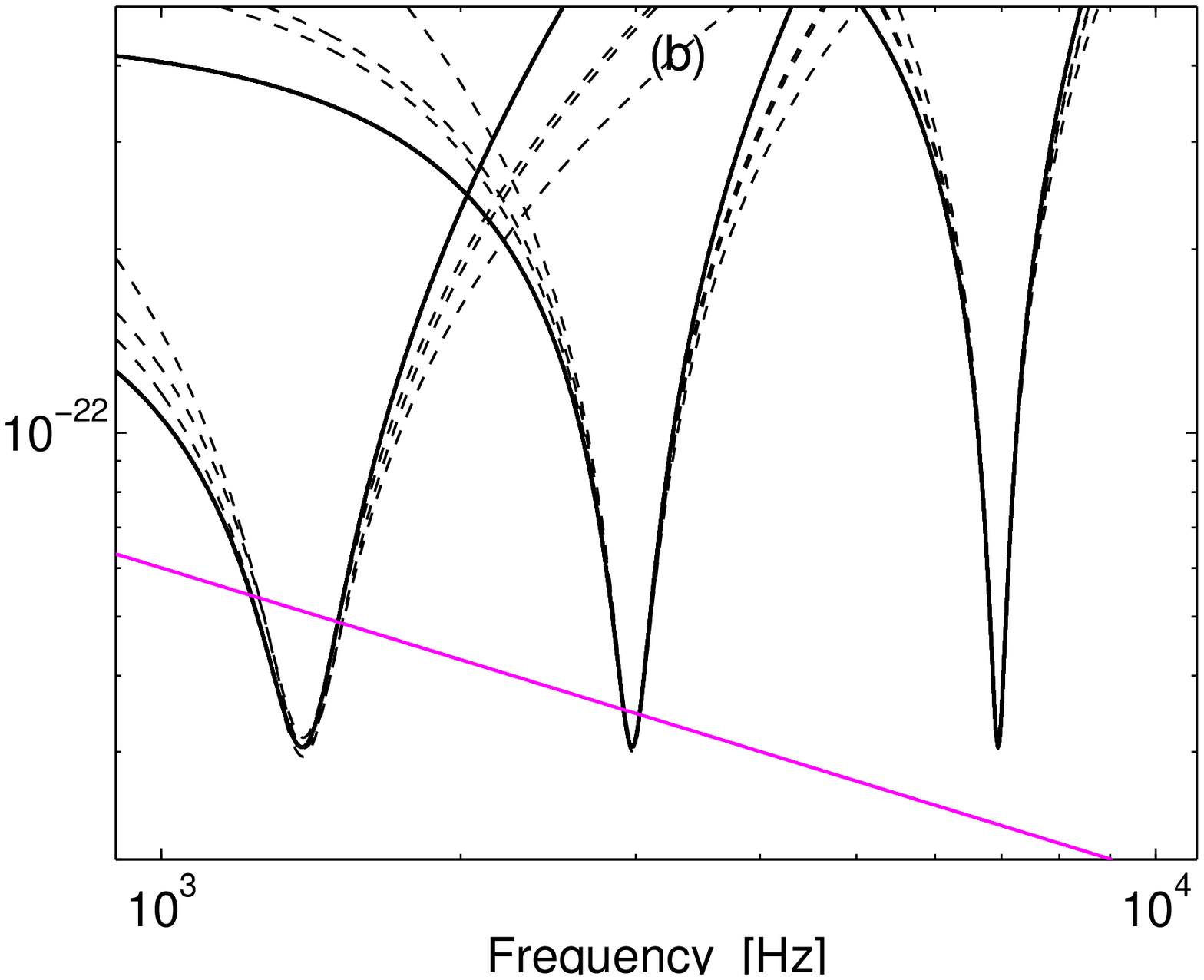}}
  \vspace{0mm}
  \caption{a) The curved lines represent squeezed light improved quantum noise
  spectral densities for four different quadrature angles.\\
           b) Zoom into the optical resonance, now for three different SR cavity detunings.}
  \label{noise3}
\end{figure}

Starting from the squeezing enhanced curve in Fig.~\ref{noise1}a,
we have plotted three additional lines (dashed) in
Fig.~\ref{noise3}a. Together with the lower solid curve they form
an array of squeezed spectral noise densities at different fixed
output quadrature detection angles $\zeta$. Here $\zeta$ is varied
from 0 (solid line) to $3/4\,\pi$ in equidistant steps. It can be
seen that above the optical resonance and at the opto-mechanical
resonance non-zero values for $\zeta$ provide lower quantum noise
floors. The lower boundary of the array is indeed achievable by
applying the variational output scheme. In Fig.~\ref{noise3}b we
zoom into the optical resonance and consider three different
detunings of the SR-cavity ($\phi=0.035, 0.075$ and
$0.175\,\rm{rad}$). It can be seen that at all three resonance
frequencies, variational output will give an improvement, but only
far above the noise minima in the wings of the resonances. This
improvement is not significant for GEO\,600s noise-floor above
1~kHz.

\section{Conclusion}
\label{CON}

We have shown that squeezed light will be able to reduce the
overall noise floor of the signal-recycled gravitational wave
interferometer GEO\,600 at frequencies above 1~kHz, where GEO\,600
is expected to be shot-noise limited. Our investigations are based
on a non-classical vacuum state of light with quantum noise
variance reduced by -6~dB at sideband frequencies between 1 and
10~kHz. If such a state is injected into the dark port of the
interferometer, then GEO\,600 with its current design parameters
can be made thermal noise limited up to 4~kHz without further
reduction of bandwidth. The interferometer might reach a strain
sensitivity of $6\times10^{-23}$ above 1~kHz (tunable) with a
bandwidth of around 350~Hz. An optimized rotation of the squeezed
quadrature along the spectrum is one of the essential properties
of the squeezed light used.
We have shown, for the special case considered here, that just one
filter cavity is sufficient to provide the desired rotation, and
can be realized from GEO\,600s signal-recycling cavity with
additional polarizing optics.\\


\appendix


\begin{thebibliography}{12}


\bibitem{Cav81} C.~M.~Caves, Phys. Rev. D {\bf 23}, 1693 (1981).

\bibitem{Unruh82} W.~G.~Unruh, in  {\it Quantum Optics, Experimental Gravitationa, and Measurement Theory}, edited by P.~Meystre and M.~O.~Scully (Plenum, New York, 1982), p.~647.

\bibitem{GLe87} J.~Geabanacloche and G.~Leuchs, J.~Mod.~Opt. {\bf 34}, 793 (1987).

\bibitem{JRe90} M.~T.~Jaekel and S.~Reynaud, Europhys.~Lett.~{\bf 13}, 301 (1990).

\bibitem{PCW93} A.~F.~Pace, M.~J.~Collett, and D.~F.~Walls, Phys. Rev. A {\bf 47}, 3173 (1993).

\bibitem{KLMTV01} H.~J.~Kimble, Y.~Levin, A.~B.~Matsko, K.~S.~Thorne, and S.~P.~Vyatchanin, Phys. Rev. D {\bf 65}, 022002 (2001).


\bibitem{CDh96} V.~Chickarmane and S.~V.~Dhurandhar, Phys. Rev. A {\bf 54}, 786 (1996).

\bibitem{CDRGBM98} V.~Chickarmane, S.~V.~Dhurandhar, T.~C.~Ralph, M.~Gray, H.-A.~Bachor, and D.~E.~McClelland, Phys. Rev. A {\bf 57}, 3898 (1998).

\bibitem{HCCFVDS03}J.~Harms, Y.~Chen, S.~Chelkowski, A.~Franzen, H.~Vahlbruch, K.~Danzmann, and
R.~Schnabel,
Phys. Rev. D {\bf 68}, 042001 (2003).

\bibitem{geo02} B.~Willke et al., Class. Quantum Grav. {\bf 19}, 1377 (2002).




\bibitem{DHKHFMW83pr} R.~W.~P.~Drever {\it et al.}\ in
{\it Quantum Optics, Experimental Gravitationa, and Measurement Theory},
edited by P.~Meystre and M.~O.~Scully (Plenum, New York, 1983), p.~503--514.


\bibitem{Mee88} B.~J.~Meers, Phys. Rev. D {\bf 38}, 2317 (1988).

\bibitem{BCh01b} A.~Buonanno and Y.~Chen, Phys. Rev. D {\bf 64}, 042006 (2001).

\bibitem{BCh01a} A.~Buonanno and Y.~Chen, Class. Quantum Grav. {\bf 18}, L95 (2001).

\bibitem{BCh02a} A.~Buonanno and Y.~Chen, Phys. Rev. D {\bf 65}, 042001 (2002).

\bibitem{BCh03a} A.~Buonanno and Y.~Chen, Phys.Rev. D {\bf 67}, 062002 (2003).

\bibitem{FHH02} C.~L.~Fryer, D.~E.~Holz, and S.~A.~Hughes, Astrophys. J. {\bf 565}, 430 (2002).



\bibitem{CSc85}C.~M.~Caves and B.~L.~Schumaker, Phys. Rev. A {\bf 31}, 3068 (1985);
B.~L.~Schumaker, and C.~M.~Caves, Phys.~Rev.~A {\bf 31}, 3093 (1985).

\bibitem{BCM03} A.~Buonanno, Y.~Chen and N.~Mavalvala, Phys.~Rev.~D {\bf 67}, 122005 (2003).










\bibitem{KSMBL02} K.~McKenzie, D.~A.~Shaddock, D.~E.~McClelland, B.~C.~Buchler, and P.~K.~Lam,
Phys. Rev. Lett. {\bf 88}, 231102 (2002).

\bibitem{SLMS98} K.~Schneider, M.~Lang, J.~Mlynek, and S.~Schiller, Opt. Ex. {\bf 2}, 59 (1998).

\bibitem{LRBMBG99} P.~K.~Lam, T.~C.~Ralph, B.~C.~Buchler, D.~E.~McClelland,
H.-A.~Bachor, and J.~Gao, J.~Opt.~B {\bf 1}, 469 (1999).

\bibitem{BSTBL02} W.~P.~Bowen, R.~Schnabel, N.~Treps, H.-A.~Bachor, and P.~K.~Lam,
J. Opt. B, 4, 421 (2002).
\bibitem{BISM98} G.~Breitenbach, F.~Illuminati, S.~Schiller, and J.~Mlynek, Europhys. Lett. {\bf 44}, 192
(1998).

\bibitem{PCh02} P.~Purdue and Y.~Chen, Phys.~Rev.~D {\bf 66} 122004 (2002).

\end{thebibliography}
\end{document}